\documentstyle[12pt]{article}

\begin{document}
\setcounter{page}{1}
\pagestyle{plain} \vspace{1cm}
\begin{center}
\Large{\bf  Strong electroweak
 phase transition in a model with extended scalar sector}\\
\small \vspace{1cm} {\bf
M. Saeedhoseini}\quad and
\quad{\bf A. Tofighi}\footnote{A.Tofighi@umz.ac.ir}\\\

\vspace{0.5cm} {\it Department of Physics, Faculty of Basic Sciences,\\
University of Mazandaran,
P. O. Box 47416-95447, Babolsar, Iran}\\

\end{center}
\vspace{1.5cm}
\begin{abstract}
In this paper we consider an extension of the Standard Model($SM$)
with additional gauge singlets which
 exhibits a strong first order phase transition. Due to this
first order phase transition in the early universe gravitational
waves are produced. We estimate the contributions such as the sound
wave, the bubble wall collision and the plasma turbulence to the
stochastic gravitational wave background,
 and we find that the strength at the peak
frequency is large enough to be detected at future gravitational
interferometers such as eLISA. Deviations in the various Higgs boson
self couplings are also evaluated.
\end{abstract}
\vspace{1.5cm}

\section{Introduction}
  The discovery of a narrow resonance, with a mass near $125$
$GeV$, at the Large Hadron Collider $(LHC)$ with properties similar
to those of the Higgs boson predicted by the Standard Model $(SM)$
$[1,2]$, sparked a lot of excitements among high energy physicists.
But in spite of this important discovery the $(SM)$ is considered to
be incomplete. For instance, in the Standard Model of particle
physics a strong first order phase transition $(SFPT)$ does not
occur $[3]$. However $SFPT$ is needed to justify the baryon
asymmetry of our universe $[4]$, moreover, the $SM$ does not have a
candidate for dark matter
($DM$).\\
Therefore, some new models are required to address these issues. A
popular model is to couple a singlet scalar to Higgs boson. In Ref.
$[5]$ it is shown that, it is possible to modify the standard theory
by adding a scalar which possesses a discrete $Z_2$ symmetry and to
address the issue of the dark matter of the universe, within the
frame work of singlet extended $SM$, the issue of dark matter has
been studied in $[6-13]$, while electroweak phase transition was
studied in Refs. $[14-23]$ and in Refs. $[24-25]$ the authors
attempt to explain electroweak phase
transition $EWPT$ and dark matter by singlet extended $SM$.\\
Another class of models are the multi-singlet extensions of the $SM$
model $[26-36]$. These models have a larger parameter space in
comparison to the singlet extended models, hence they can address
several issues, in Refs. $[26, 27]$ cosmological implications of
such models with classical conformal invariance is presented.
Electroweak phase transitions in two-Higgs doublet model is analyzed
in $[37, 38]$ and within supersymmetric models in $[39-45]$.
 A comprehensive review of $EWPT$ within
various models has been given in
$[46]$.\\
In order to investigate the dynamics of the electroweak phase
transition $EWPT$ one has to utilize techniques from the domain of
thermal field theory $[47-50]$. The occurrence of a first order
phase transition, requires that the electroweak breaking and
preserving minima to be degenerate, an event which happens at a
critical temperature $T_c$. Moreover, to prevent the washout of any
baryon asymmetry by electroweak sphalerons, the electroweak phase
transition
 must be strongly first
order. Namely the ratio of vacuum expectation value of the Higgs
field to the critical temperature needs to be greatar than unity. As
we describe in the next section if a first order phase transition
occurs in the early universe, the dynamics of bubble collision  and
subsequent turbulence of the plasma are expected to generate
gravitational wave ($GW$). If we detect these $GW$ then we can
obtain information about symmetry breaking in early universe. $GW$
signals from phase transitions has been discussed in
$[51-59]$.\\
In $[60]$ the authors study $EWPT$ within several exotic models and
in $[61]$ an analysis of the $EWPT$ of a large number of minimal
extensions of the $SM$
 and their classically conformal limits is presented.
Complex conformal singlet extension of $SM$ with emphasis on
 the issue of dark
matter and Higgs phenomenology is studied in $[62]$. Recently
baryogenesis within a $\varphi^6$ model is addressed in $[63]$. An
investigation of  electroweak phase transitions in a singlet
extended model in the $100$ ($TeV)$ range is given in $[64]$. The
authors of $[65]$ study strong first order $EWPT$ in a singlet
scalar extension of the $SM$ where the singlet scalar is coupled
non-minimally to gravity. In this scheme the singlet field first
derives inflation and at a later time causes a strong $EWPT$, in a
new study a first order $EWPT$ in the $SM$ is
obtained by varying Yukawas during phase transition $[66]$.\\
In this work we propose a new model and we investigate the strength
of $EWPT$ within this model. In this model which is a generalization
of $[19]$, $N$ real gauge singlet are coupled to Higgs boson via
trilinear interactions. Previous studies of multi-scalar singlet
extension of $SM$ impose separate $Z_2$ symmetries on the singlets
$[26, 28, 29, 31, 33, 35, 36]$, however in our model we do not
require such symmetry. In spite of it's simple form , this model has
a very rich phenomenology. The main feature of the singlet extended
$SM$ model (without $Z_2$ symmetry) is that the potential barrier
between the true and the false vacua necessary for first order
$EWPT$ can be formed mainly by tree-level interactions. But in the
singlet extended $SM$ (with $Z_2$ symmetry) non-decoupling loop
effects are needed for the occurrence of a strong first order
$EWPT$, however, these models have $DM$ candidate. In Ref. $[67]$ a
non-minimal composite model based on the coset $\frac{SO(7)}{SO(6)}$
has been considered. At low energy the scalar sector of their model
is composed of two scalars, one with an unbroken $Z_2$ symmetry and
another scalar with a broken $Z_2$
symmetry.\\
The plan of this paper is as follows:\\
In section two we summarize the basic notions of $EWPT$. We describe
the finite temperature effective potential at one-loop. Then by
emphasizing the underlying physical mechanisms, we describe the
basic quantities of interest such as the strength of a phase
transition, the rate of variation of bubble nucleation rate per
volume and the ratio of the latent heat released at the phase
transition to the radiation energy density. In section three we
consider
 a simple extension of the $SM$ by the addition of $N$
real scalar gauge singlets with trilinear coupling to Higgs. We
present the phenomenology of the model for $N=2$ and due to lack of
protective symmetry these gauge singlets are not candidates for $DM$
and we discuss the issue of observability of gravitational wave of
our model. And finally in section four we present our
conclusions. Technical details are explained in the appendix.\\

\section{Electroweak Phase Transition}
In this section we summarize basic notions and definitions of the
electroweak phase transitions.\\
 The observable universe consists predominantly of matter.
The asymmetry between the matter and anti-matter content of the
universe is expressed by the baryon to photon ratio
\begin{equation}
\rho=\frac{n_b-n_{\bar{b}}}{n_{\gamma}}\sim 10^{-9},
\end{equation}
where $n_b$, $n_{\bar{b}}$ and $n_{\gamma}$ are the number densities
of baryons, antibaryons and photons. In a symmetric universe one
expects $\rho=0$ but experiments reveal that $\rho$ has a tiny but
non-zero value. This paradox can be resolved by requiring baryon
number violation, C and CP violation and departure from thermal
equilibrium $[4]$.\\
Baryogenesis is the physical process which is responsible for this
observed baryon asymmetry of universe (BAU). In electroweak
baryogenesis one assumes that the physical mechanism is the
occurrence of a strong first order electroweak phase transition
(EWPT) namely a smooth or a weak phase
transition can not explain BAU $[68, 69]$.\\
A convenient tool for investigation of EWPT is effective potential.
For any quantum field theory if we replace the quantum field by it's
vacuum expectation value in the presence of a source the result for
the potential energy to lowest order in perturbation theory is
called the effective potential. The physical meaning of the
effective potential is that it represent an energy density.
 In
general effective potential contains other terms (loop corrections)
$[70, 71]$. By using path integrals it is possible to find effective
potential. In an Euclidean space time at one loop order the result
is
\begin{equation}
V^{eff}(\varphi_c)=V_0(\varphi_c)+\frac{i}{2}\int\frac{d^4p}{(2\pi)^4}
log[\frac{p^2+m^2(\varphi_c)}{p^2}]+O(\hbar^2)+...,
\end{equation}
which is known as the Coleman-Weinberg potential.
The first term in $eq.(2)$ is the classical tree-level potential.\\
 While considering physical processes in a hot environment such as
early universe a proper method is finite temperature field theory
$[47-50]$. The expression for the one loop effective potential at
finite temperature is $[72]$
\begin{equation}
V^{eff}_1(\varphi_c,T)=\sum_i\frac{n_iT^4}{2\pi^2}J_{\mp}(\frac{m_i(\varphi_c)}{T}),
\end{equation}
where $n_i$ is the number of degrees of freedom of the particle,
$m_i$ is the field dependent particle mass, and
\begin{equation}
J_{\mp}=\pm\int^\infty_0 dy y^2\log [1\mp exp(-\sqrt{x^2+y^2})],
\end{equation}
and $J_-(x)$, $J_+(x)$ denotes the contribution from bosons,
fermions. In addition there is another contribution to thermal
effective potential
 from the daisy
subtraction $[73, 74]$.
\\
The best way for the evaluation of the finite temperature effective
 potential is to use
special
packages (codes) developed for a specific model.\\
But when the temperature is much greater than the masses of the
various particles of the system under considerations it is possible
to expand the $J_\mp$ in a power series of $\frac{m}{T}$. In this
work we use this high temperature approximation. Recently, a new
scheme for computation and resummation of thermal masses beyond the
high-temperature approximation in general beyond SM scenarios has
been proposed $[75]$.\\
Let us consider the shape of the effective potential. For a generic
model,
 as the
universe cooled down at temperature above a critical temperature
$T_c$ the effective potential had an absolute minima which was
located at the origin. At $T_c$ there were two degenerate minima,
which were separated by an energy barrier. At temperature below the
critical temperature the second minimum became the global one and
presently there is no energy barrier. Moreover the rate of expansion
of the universe slowed, as the Hubble parameter $H$ which
characterizes the rate of expansion of the universe depends
quadratically on the temperature.\\
In $1976$ 't Hooft discovered that baryon number is violated in the
Standard Model and it is due to transition between two topologically
distinct $SU(2)_L$ ground states $[76, 77]$. While at zero
temperature, the probability for barrier penetration is vanishingly
small, at non-zero temperature the transition between two ground
state differing by a unit of topological charge can be achieved by a
classical motion over the barrier. Unstable static solutions of the
field equations with energy equal to the height of the barrier
separating two topologically distinct $SU(2)_L$ ground states
(sphalerons) have been reported in $[78]$. Hence in the electroweak
baryogenesis , the baryon asymmetry of universe is generated through
the sphaleron process in the symmetric phase during the $EWPT$.
 At the symmetric phase the rate of
baryon violating process which we denote by $\tilde{\Gamma}^{sph}$
is a quartic function of the temperature, hence in this phase
$\tilde{\Gamma}^{sph}>> H$.\\
But the third conditions to generate BAU is the departure from
thermal equilibrium. Hence, the baryon number changing sphaleron
interaction must quickly decouple in the broken phase, that is
$\tilde{\Gamma}^{sph}< H$. But the rate of sphaleron induced
 baryon violating process is suppressed by a Boltzman factor
\begin{equation}
\tilde{\Gamma}^{sph}\propto exp[-\frac{E_{sph}(T)}{T}],
\end{equation}
but at phase transition $E_{sph}(T)\propto \varphi_c$
\begin{equation}
\frac{\varphi_c}{T_c}\geq 1,
\end{equation}
where $\varphi_c$ is the broken phase minimum at the critical
temperature $T_c$.\\
 A first order phase transition proceeds by nucleation of bubbles of
the broken symmetry phase within the symmetric phase. The underlying
mechanisms for bubble nucleation are quantum tunneling and thermal
fluctuations. These bubbles then expand, merge and collide.
Gravitational waves are produced due to collision of bubble walls
and turbulence in the plasma after the collisions. In addition while
these bubbles pass through the plasma sound waves are created. These
sound waves can provide additional sources of gravitational waves.\\
A crucial parameter for the calculation of the gravitational wave
spectrum is the rate of variation of the bubble nucleation rate per
volume, called $\beta$. It is common to use a normalized
dimensionless parameter which is defined as
\begin{equation}
\tilde{\beta}=\frac{\beta}{H_*},
\end{equation}
where, $H_*$ denotes the Hubble parameter at the time of phase transition.\\
 The
density of latent heat released into the plasma is
\begin{equation}
\epsilon_*=[-V^{eff}_{min}(T)+T\frac{d}{dT}V^{eff}_{min}(T)]_{T=T_*},
\end{equation}
where $V^{eff}_{min}(T)$ is the temperature-dependent true minimum
of the effective potential of the scalar fields which causes the
phase transition moreover, it's value must be set to zero by adding
a constant at each time. Another dimensionless parameter for
characterizing the spectrum of gravitational wave is
\begin{equation}
\alpha=\frac{\epsilon_*}{\rho_{rad}},
\end{equation}
 where the radiation energy
density of the plasma
 $\rho_{rad}=\frac{\pi^2}{30}g_* T_c^4$. And the parameter
$g_*$ is the effective degrees of freedom in the thermal bath at the
phase transition. In this work we assume $g_*= 106.75 + N_S$ where
$N_S$ denotes the number of  singlet scalars that facilitates the
electroweak phase transition.\\

\section{ The Model }
In Ref. $[19]$  the effects of a light scalar on the electroweak
phase transition has been considered. In their model the scalar
sector has been extended by an addition of a singlet, which has
 a trilinear interaction with the Higgs boson. Recently an extension of the $SM$
with addition of $N$ isospin-singlet, which has a quartic
interaction with the Higgs boson has been considered $[57]$. Here we
consider a generalization of model of $[19]$.
 At zero temperature the effective potential of the scalar sector of our model is\\
\begin{equation}
V_0=-DT^2_0\varphi^2+\frac{\lambda}{4}\varphi^4+\frac{1}{2}\sum^N_{i=1}m^2_{i}s_i^2
+\varphi^2\sum^N_{i=1}\kappa_i{s_i}.
\end{equation}
Since we have not included quartic self coupling for the extra
scalars , in order to have a stable potential , we assume that the
squared of the mass parameters
 of the extra
scalars ($m^2_i$)are positive definite (See appendix for detail).\\
At high temperature we have
\begin{equation}
V_T=D(T^2-T^2_0)\varphi^2-ET\varphi^3+\frac{\lambda_T}{4}\varphi^4
+\frac{1}{2}\sum^N_{i=1}m^2_{i}s_i^2
+\varphi^2\sum^N_{i=1}\kappa_i{s_i}+\frac{T^2}{12}\sum^N_{i=1}\kappa_is_i.
\end{equation}
The parameters of $eq.(11)$ are given by\\
\begin{eqnarray}
D&=&\frac{1}{8 v^2}(2m^2_W+m^2_Z+2m^2_t+2\lambda v^2)
\nonumber\\
E&=&\frac{1}{8\pi v^3}(4m^3_W+2m^3_Z)
\nonumber\\
\lambda_T&=&\lambda-\frac{1}{16\pi^2 v^4}(6m^4_Wln\frac{m^2_w}{a_BT^2}
+3m^4_Zln\frac{m^2_Z}{a_BT^2}-12m^4_tln\frac{m^2_t}{a_FT^2})
\nonumber\\
ln a_B&=&3.91, \qquad ln a_F=1.14.
\end{eqnarray}
The critical temperature of model is given by
\begin{equation}
T_c=\frac{T_0}{\sqrt{1-\frac{E^2}{D(\lambda_T-2\zeta)}-\frac{\zeta}{12D}}},\qquad
where\qquad \zeta=\sum^N_{i=1}\frac{\kappa^2_i}{m^2_i}.
\end{equation}
The strength of the phase transition is denoted by $\xi$ and it is
given by
\begin{equation}
\xi=\frac{\varphi_c}{T_c}=\frac{2E}{\lambda_T-2\zeta}.
\end{equation}
Moreover, $s_{ic}$ is the  $vev$ of the scalar field
$s_i$ at the second minimum
of the effective potential at $T_c$ is
\begin{equation}
s_{ic}=-\frac{\kappa_i}{m^2_i}(\varphi_c^2+\frac{T^2_c}{12}).
\end{equation}
\\
\subsection{Phenomenology of the models with N trilinear interactions}
The structure of the scalar mass matrix is\\
$\left(%
\begin{array}{cccccc}
  2\lambda v^2 & 2\kappa_1 v & 2\kappa_2 v & 2\kappa_3 v & ... & ... \\
   2\kappa_1 v & m^2_1 & 0 & 0 & 0 & 0 \\
   2\kappa_2 v& 0 &m^2_2  & 0 & 0 & 0 \\
   2\kappa_3 v&  & 0 & m^2_3 &...  &...  \\
  ... & .. &...  & ... & ... & ... \\
  ... & ... & ... &...  &...  &...  \\
\end{array}%
\right).$\\
The physical mass squared of the scalar sector of our model are the eigenvalues of this matrix. The characteristic
equation is\\
$det\left(%
\begin{array}{cccccc}
 \omega- 2\lambda v^2 & -2\kappa_1 v & -2\kappa_2 v & -2\kappa_3 v & ... & ... \\
   -2\kappa_1 v &\omega- m^2_1 & 0 & 0 & 0 & 0 \\
   -2\kappa_2 v& 0 &\omega-m^2_2  & 0 & 0 & 0 \\
   -2\kappa_3 v&  & 0 & \omega-m^2_3 &...  &...  \\
  ... & .. &...  & ... & ... & ... \\
  ... & ... & ... &...  &...  &...  \\
\end{array}%
\right)=0.$\\
Hence in general one should solve a polynomial of degree $(N+1)$ of
$\omega$. From the above equation we see that the coefficient of
$\omega^{N+1}$ is unity. If we designate the coefficient of
$\omega^{N}$ by $\alpha$, then $-\alpha$ is equal to the sum of
physical masses of the model, namely
\begin{equation}
2\lambda v^2+\sum^N_{i=1} m^2_i=m_H^2+\sum^N_{i=1} \chi^2_i
\end{equation}
where the physical mass of the $i th$ scalar is denoted by $\chi_i$.
From this relation we find an important relation about the Higgs
self coupling, namely
\begin{equation}
\lambda -\lambda_{SM}=\frac{1}{2v^2}\sum^N_{i=1}(\chi^2_i- m^2_i),
\end{equation}
where $\lambda_{SM}$ is the Higgs quartic self coupling of the standard model.\\
We see that if the mass parameters of the scalars as well as the
physical masses of the scalars are much smaller than $v$ then the
the deviation of the parameter $\lambda$ from the
Higgs self-coupling of the standard model will be very small.\\

\subsection{The special case N=2}
Here we consider the case $N=2$. The scalar mass matrix of the
model at zero temperature for this case is.\\

$\textsl{$M^2$}=
\left(%
 \begin{array}{ccc}
  2\lambda v^2 & 2\kappa_1 v& 2\kappa_2 v \\
  2\kappa_1 v & m^2_1 & 0 \\
   2\kappa_2 v& 0 & m^2_2 \\
\end{array}%
\right).$\\
The physical mass squared of the model can be obtained from
\begin{equation}
\omega^3+A\omega^2+B\omega+C=0,
\end{equation}
where
\begin{eqnarray}
 A&=&-(2\lambda v^2+m^2_1+m^2_2),\qquad
 B=2v^2[\lambda(m^2_1+m^2_2)-2(\kappa_1^2+
\kappa_2^2)]+m^2_1m^2_2\nonumber\\
C&=&4v^2(\kappa^2_1m^2_2+\kappa^2_2m^2_1)-2\lambda^2v^2m_1^2m_2^2.
\end{eqnarray}
Now one of the eigenvalues is equal to $m^2_H$, hence we obtain
\begin{equation}
m^6_H+Am_H^4+Bm_H^2+C=0,
\end{equation}
Moreover,by minimizing the potential with respect to variables $s_1, s_2$ we obtain
\begin{equation}
s_i=-\frac{\kappa_iv^2}{m^2_i},\qquad i=1,2.
\end{equation}
By minimizing the potential with respect to variable $\varphi$ we obtain
\begin{equation}
-2DT^2_0+\lambda v^2-2\sum^2_{i=1}\frac{\kappa_i^2v^2}{m^2_i}=0.
\end{equation}
The parameters of the model must satisfy eqs.(20,22). Hence the
parameter space of the model in this case contains
four independent parameters ($\kappa_1$,$\kappa_2$, $m_1$,$m_2$).\\
 If we subtract
eq.(20) from eq.(18) we get
\begin{equation}
\chi^4+(A+m_H^2)\chi^2+m_H^4+m_H^2 A+B=0,
\end{equation}
Therefore,the physical mass squared of the singlets are
determined.\\
But models with extended Higgs sectors predicting strongly first
order phase transition simultaneously predict a significant
deviation in the triple Higgs boson coupling as well $[79]$. This
deviation at the tree level in $[17]$ and at loop level in $[33,79]$
has been studied, with
\begin{equation}
\Delta_{hhh}=\frac{\lambda^{MSM}_{hhh}-\lambda^{SM}_{hhh}}{\lambda^{SM}_{hhh}},
\end{equation}
where $\lambda^{SM}_{hhh}$ is the Higgs triple coupling of the $SM$
and $\lambda^{MSM}_{hhh}$ is the Higgs triple coupling of the
multi-singlet extension of the $SM$. Collider experiments could
measure the Higgs triple coupling.
 Here we want to explore the region of intermediate mass of the
singlets.
 But there are bounds on Higgs-Portal models from the LHC Higgs
 data $[80, 81, 82]$. For instance the Higgs doublet is mixed with the
 extra singlet scalars and the mixing element $cos(\varphi)$
between the CP-even component of the doublet ($\varphi$) and the
physical Higgs (whose $m_H$=125.09 GeV) is not arbitrary and it is
subject of a constraint coming from the Higgs coupling to the W
gauge bosons. Current data $[81]$ suggests $\cos(\varphi)> 0.86$.\\
The results are presented in Table $1$ for the onset of a strong
$EWPT$, namely $\xi=1$. For each configuration in the table we
present our results for the deviation of Higgs triple coupling as
well. Hence by adding one scalar to the model we can have singlets
in the intermediate mass
region.\\\\
\clearpage
\begin{tabular}{|c|c|c|c|c|c|c|c|c|c|}
  \hline
  Set& $\kappa_1$ & $\kappa_2$ & $m_1(GeV)$ & $m_2(GeV)$ &$\chi_1(GeV)$
  &$\chi_2(GeV)$&$cos(\varphi)$&
  $\Delta\lambda_{hhh}$&$\Delta\lambda_{hhhh}$\\
  \hline
  I & 0.2 & 3.7 & 69.2 & 16.0 & 6.2 & 69.2 & 0.993&13.9 \%&4.1\%\\
  II & 0.5 & 3.1 & 18.7 & 13.5 & 5.2 & 18.6 &0.995&14.5 \%&3.0\%\\
  III & 0.8& 2.3 & 11.2 & 10.4  & 4.0 & 11.1& 0.997& 15.1 \%&1.8\%\\
  IV & 1.2 &0.7& 18.1 & 3.2 & 17.5& 1.3 & 0.999 & 5.7 \%&0.6\%\\
  \hline
\end{tabular}\\\\

Table $1$: Different configurations associated with the onset of a
strong
first order phase transition ($\xi=1$).\\\\

For completeness the predictions of the model for the deviation of
Higgs boson quartic coupling $\Delta\lambda_{hhhh}$ are given in
Table $1$.\\
 Moreover, the existence of the extra singlet scalars could affect
the total Higgs decay if they are light enough, which becomes
\begin{equation}
\Gamma_{total}(h)=cos^2(\varphi)\Gamma_{total}(h_{SM}),
+\sum\Gamma(h\rightarrow s_i+s_k)
\end{equation}
where $s$ denote all the scalars. The deviation from the $SM$ value
$\Delta\Gamma_{total}< 1.4 (MeV)$.\\
In the $SM$, a total Higgs decay width around 4 $MeV$ is predicted.
In this work we assume the decay  $h\rightarrow s_i+s_k$ is an
invisible decay. However, current analysis $[82]$ suggests that the
Higgs invisible decay branching ratio should be less than $17\%$. In
Table $2$ we present Higgs invisible decay branching ratio in
various decay modes, as well the total Higgs invisible branching
ratio for all of Higgs invisible decay modes. The invisible Higgs
width , total Higgs width and the deviation of total Higgs width of
our model from that of the $SM$ are also shown. The unit for the
variuos width in this Table is ($MeV$), moreover in our calculation
we have assumed
$\Gamma_{total}(h_{SM})=4 (MeV)$.\\
It turns out that the critical temperature for this model $T_c< 100
(GeV)$. Hence it is a good approximation to use $\lambda$ instead of
$\lambda_T$ and in Ref. $[19]$ this approximation
 is used to study $SFPT$ for the case $N=1$ but
the mass of the light scalar is up to $12$ $GeV$, in Ref. $[20]$ a
one loop study of the same model has been presented
 but the mass of the light
scalar to catalyze a $SFPT$ is up to $20$ $GeV$. But for the model
presented in this work and using this approximation for the case
$N=2$ the mass of the scalar to catalyze a $SFPT$ is $69$ $GeV$.
By adding more scalars we expect to have heavy singlets.\\
\clearpage

\begin{tabular}{|c|c|c|c|c|c|c|c|}
  \hline
  Set & $B(h\rightarrow s_1s_1)$&$B(h\rightarrow s_1s_2)$&
  $B(h\rightarrow s_2s_2)$&$B_{tot}(h\rightarrow inv)$
  & $\Gamma_{inv}(h)$& $\Gamma_{tot}(h)$ & $\Delta\Gamma_{tot}$ \\
  \hline
  I & 1.45 \%& 0.31 \%& 0.0 \%& 1.49 \%& 0.06 & 4.006 & 0.006 \\
  II & 0.75\% & 0.04 \%& $3.6\times 10^{-5}$ \%& 0.8 \%& 0.03 & 3.992 & 0.008 \\
  III & 0.24 \% & 0.004 \%& $1.3\times10^{-6}$ \%& .24 \%& 0.01 & 3.99& 0.01 \\
  IV & 0.02 \%& $4.6\times10^{-7}$ \%& $6.3\times10^{-12}$ \%& 0.02 \% & $5.8\times10^{-4}$ & 3.993& 0.007 \\
  \hline
\end{tabular}\\\\

Table $2$: Values of Higgs invisible branching ratios for various
decay modes, total value of Higgs invisible branching ratio, Higgs
invisible width, Higgs total width and deviation of the Higgs width
from the SM value are shown for various configurations. The unit for
width in this table is $MeV$. All four configurations are
 consistent
with current
experimental data.\\\\

\subsection{Detection of gravitational waves} The direct detection of
gravitational waves by LIGO $[83]$ generated a lot of interest among
researchers in  cosmology, astrophysics and particle physics. Major
sources of gravitational waves are inflation,
compact binary systems or cosmological phase transitions.\\
In this section under phenomenological description we aim to make an
estimate for the observability of the gravitational waves produced
by the model presented in this work.\\
In all of previous multi-singlet models a $Z_2$ symmetry (either
broken or unbroken) is imposed on the fields. However, neither of
the two fields of our model has this symmetry. This is the main
difference between our model and previous studies.\\
The main parameters that are of value for obtaining the spectrum of
gravitational waves are the parameter $\alpha$ and $\beta$, which we
discussed in section two. And they are obtained from the thermal
effective potential.\\
For the set $I$ of Table $1$ and from $eqs. (8,9)$ we obtain
$\alpha=0.16$. But the velocity of the bubble wall and the
efficiency factor, the fraction of the latent heat which is
converted to the kinetic energy of the plasma are determined from
\begin{equation}
v_b=\frac{\frac{1}{\sqrt{3}}+\sqrt{\alpha^2+\frac{2\alpha}{3}}}{1+\alpha},\qquad
\kappa=\frac{1}{1+0.715 \alpha}(0.715
\alpha+\frac{4}{27}\sqrt{\frac{3\alpha}{2}}),
\end{equation}
Thus for this configuration $v_b=0.81$ and $\kappa=0.17$. The
 The peak frequency for bubble collision contribution is given by
 $[84]$
\begin{equation}
f_{col}=16.5\times
10^{-6}\frac{0.62}{v^2_b-0.1v_b+1.8}\frac{\beta}{H_*}\frac{T_*}{100}(\frac{g_*}{100})^{\frac{1}{6}}.
\qquad Hz
\end{equation}
And the energy density at this frequency is given by
\begin{equation}
\Omega
h^2_{col}=1.67\times10^{-5}(\frac{\beta}{H_*})^2\frac{0.11v_b^3}{0.42+v^2_b}
(\frac{\kappa \alpha}{1+\alpha})^2 (\frac{g_*}{100})^{\frac{-1}{3}}
\end{equation}
where $h$ is the reduced Hubble constant at present.\\
Another source of gravitational waves is the compression waves in
the plasma (sound waves) and the peak frequency is
\begin{equation}
f_{sw}=1.9\times
10^{-5}\frac{\beta}{H_*}v^{-1}_b\frac{T_*}{100}(\frac{g_*}{100})^{\frac{1}{6}},
\qquad Hz
\end{equation}
the energy density at this peak frequency is obtained from
\begin{equation}
\Omega h^2_{sw}=2.65\times10^{-6}(\frac{\beta}{H_*})^{-1}
(\frac{\kappa \alpha}{1+\alpha})^2
(\frac{g_*}{100})^{\frac{-1}{3}}v_b.
\end{equation}
And finally the
 peak frequency for gravitational waves which are caused by the
 turbulence of the plasma is
\begin{equation}
f_{turbo}=2.7\times
10^{-5}\frac{\beta}{H_*}v^{-1}_b\frac{T_*}{100}(\frac{g_*}{100})^{\frac{1}{6}}\qquad
Hz,
\end{equation}
And the peak energy density of this part is given by
\begin{equation}
\Omega h^2_{turbo}=3.35\times10^{-4}(\frac{\beta}{H_*})^{-1}
(\frac{\varepsilon \kappa \alpha}{1+\alpha})^{\frac{3}{2}}
(\frac{g_*}{100})^{\frac{-1}{3}}v_b
\frac{1}{2^{\frac{11}{3}}(1+8\pi\frac{f_{turbo}}{H_*})}
\end{equation}
where $\varepsilon$ denotes the fraction of latent heat that is
transformed into turbulent motion of the plasma.  We choose
$\varepsilon=0.05$.\\
Hence by knowing the values of  the parameters $\alpha$,
$\tilde{\beta}$ and $T_*$ we can obtain the spectra of the $GW$.
Even though the nucleation temperature $T_*$ is lower than the
critical temperature in this work we assume $T_*\approx T_c$.\\
The standard method of calculation of the parameter $\tilde{\beta}$
from the effective potential is to compute the Euclidean action of
the model and it is explained in $[36, 85]$, but in an approximate
scheme it is found that $[59]$,
\begin{equation}
\tilde{\beta}\approx 170-4\ln(\frac{T_*}{1GeV})-2\ln g_*,
\end{equation}
and in our case when the phase transition happens at weak scale,
$\tilde{\beta}\approx 144$.\\
Hence, in order to assess the implications of the model on the
spectra of $GW$ we present results by varying this parameter in the
interval $50\leq \tilde{\beta}\leq 250$.
In Table $3$ we present our results. The unit for various frequencies is $mHz$. \\

\begin{tabular}{|c|c|c|c|c|c|c|c|}
  \hline
   $GW Spectra$ & $f_{sw}$ & $f_{col}$& $f_{turb}$ & $\Omega h^2_{sw}$ &$\Omega h^2_{col}$
     &$\Omega h^2_{turb}$  &  $\Omega h^2_{total}$ \\
  \hline
  $\tilde{\beta}=50$ & 0.65 & 0.12 & 1.85 & $2.34\times10^{-11}$ &
  $1.98\times10^{-13}$
   & $1.67\times10^{-15}$ & $2.36\times10^{-11}$ \\
  $\tilde{\beta}=100$ & 1.30 & 0.24 & 3.69 & $1.17\times10^{-11}$ & $9.9\times10^{-14}$
  & $8.35\times 10^{-16}$& $1.18\times10^{-11}$ \\
  $\tilde{\beta}=250$& 3.25 & 0.60 & 9.23 & $4.67\times10^{-12}$
  & $3.96\times10^{-14}$ & $3.34\times10^{-16}$ & $4.71\times10^{-12}$ \\
  \hline
\end{tabular}\\\\

Table $3$: The spectra of gravitational wave as predicted by our
model. The emitted
$GW$ are within the reach of eLISA C1.\\

 The results of Table $3$ shows that in our model the contribution from
the turbulent motion of always a few orders of magnitude smaller
than the previous two. Moreover, the contribution from the sound
waves is the dominant source of the total $GW$ spectrum. By studying
the frequency dependent spectra $[84]$, we find that the peak of
energy density of the sound wave contribution and the peak of energy
density due to collision are well separated. A desirable feature
while detecting these waves.  These waves are within the reach of
future gravitational wave interferometers (eLISA C1).\\

\section{Conclusions}
 In this work we have presented a new extension of the $SM$.
In this model we amend the $SM$ by $N$ gauge singlets. In order to
avoid proliferation of the parameters we considered the most
economical model. For each scalar we allowed a mass term with a
positive squared mass parameter to insure vacuum stability in the
direction of that scalar and a triple coupling with the Higgs field
to facilitate a strong electroweak phase transition,
 and for the
special case $N=2$ we find $SFPT$ with gauge singlets in the
intermediate mass range. We also investigated the deviations of the
Higgs coupling constants from the $SM$ values. And we find that the
deviation of the triple Higgs boson coupling can be as large
$15\%$. \\
 Finally, we have obtained the gravitational wave spectrum
from the electroweak phase transition. We have shown that the
gravitational wave signal can be detected by eLISA. The present
model has a large parameter space and as a result a richer
phenomenology in comparison to $[19]$. It would be of interest to
amend the model by inclusion of quartic self-coupling of the scalars
as well as the mixing term
between scalars(see appendix for detail). \\
 It would be of interest to consider
higher values of $N$ for the model proposed in this work. These and
other related issues are presently under considerations.\\

{\bf Acknowledgements}:\\

We would like to thank the referees for their valuable comments and suggestions.\\\\

{\bf Appendix :} {\bf
Vacuum stability conditions}\\\\

 For the special case $N=2$ of the model
 the weak condition for vacuum stability is that if the value of
the fields tend to infinity then the tree-level potential is not
unbounded-from-below directions. This leads to $\lambda>0$. But if
only the field $s_1$ tend to infinity, then the condition vacuum
stability will be $m_1^2>0$ and by similar argument for the field
$s_2$ the restriction will be $m_2^2>0$.\\
However, the strong condition for vacuum stability as stated in
$[35, 86, 87]$is that the masses of the particles of a
particular model be positive.\\
The necessary conditions for a symmetric matrix A of order $3$ to
have real eigenvalues are;
\begin{eqnarray}
&&a_{11}>0,a_{22}>0,a_{33}>0,
\nonumber\\
&&\bar{a}_{12}=a_{12}+\sqrt{a_{11}a_{22}}>0,
\nonumber\\
&&\bar{a}_{13}=a_{13}+\sqrt{a_{11}a_{33}}>0,
\nonumber\\
&&\bar{a}_{23}=a_{23}+\sqrt{a_{22}a_{3
3}}>0,
\end{eqnarray}
and
\begin{equation}
\sqrt{a_{11}a_{22}a_{33}}+a_{12}\sqrt{a_{33}}+a_{13}\sqrt{a_{22}}+
a_{23}\sqrt{a_{11}}+\sqrt{2\bar{a}_{12}\bar{a}_{13}\bar{a}_{23}}>0.
\end{equation}
The constraints stated in eq. (34) applies to quantities with
dimension of squared mass. From our mass matrix of subsection $3-2$,
the first constraint is $2\lambda v^2>0$, which leads to $\lambda
>0$. Hence,
 for the model presented in this work, the criteria which provide
the necessary and sufficient vacuum stability conditions are given
by,
\begin{eqnarray}
&&\lambda>0,\qquad m^2_{1}>0,\qquad m^2_{2}>0,
\nonumber\\
&&\kappa_1>-\sqrt{\frac{\lambda m^2_1}{2}},\qquad
\kappa_2>-\sqrt{\frac{\lambda m^2_2}{2}},
\end{eqnarray}
and
\begin{equation}
\sqrt{\lambda m_1^2m_2^2}+\kappa_1\sqrt{2 m^2_2}+\kappa_2\sqrt{2
m^2_1}+\sqrt{(2\kappa_1+\sqrt{2\lambda
m_1^2)}(2\kappa_2+\sqrt{2\lambda m_2^2)}\sqrt{m_1^2m_2^2}}>0,
\end{equation}
where the conditions stated in the first line of $eq. (36)$ has been
obtained by using weak conditions for vacuum stability.\\
In our model we have not included quartic terms for the gauge
singles such as
\begin{equation}
V_{quartic}=\sum^2_{i=1}\frac{\lambda_is_i^4}{4} +\delta s_1^2s_2^2
\end{equation}
in the effective potential, where $\lambda_1$, $\lambda_2$ and
$\delta$ are dimensionless coupling parameters,
 in this case the vacuum stability
in the direction of the extra scalars is maintained if
\begin{equation}
\lambda_1>0,\qquad\lambda_2>0, \qquad
\sqrt{\lambda_1\lambda_2}>-2\delta,
\end{equation}
and the squared of the mass parameters $m^2_1$ and $m^2_2$ in
principle can assume any value (positive, zero or negative), for
instance in the multi-scalar model described in $[26]$ all of the
extra scalars do not have a mass term,
 while
in the absence of quartic terms the squared mass parameters of the
model presented in section three must be positive as required by
vacuum
stability.\\


\clearpage
\begin{thebibliography}{14}
\bibitem{1}
 ATLAS Collaboration, G. Aad et al., Phys. Lett. B {\bf 716} (2012) 1.
\bibitem{2}
CMS Collaboration, S. Chatrchyan et al., Phys. Lett. B {\bf 716}
(2012) 30.
\bibitem{3}
G. W. Anderson and L. J. Hall, Phys. Rev. D {\bf 45} (1992) 2685.
\bibitem{4}
A. Sakharov, Pisma Zh. Eksp. Teor. Fiz. {\bf 5} (1967) 32.
\bibitem{5}
V. Silveira and A. Zee, Phys. Lett. B {\bf 161}  (1985) 136.
\bibitem{6}
J. McDonald, Phys. Rev. D {\bf 50 } (1994) 3637.
\bibitem{7}
C. P. Burgess, M. Pospelov and T. ter Veldhuis, Nucl. Phys. B {\bf
619} (2001) 709.
\bibitem{8}
J. J. van der Bij, Phys. Lett. B {\bf 636} (2006) 56.
\bibitem{9}
V. Barger, P. Langacker, M. McCaskey, M. J. Ramsy-Musolf and G.
Shaughnessy, Phys. Rev. D {\bf 77} (2008) 035005.
\bibitem{10}
W. -L. Guo, Y. -L. Wu, JHEP {\bf1010}  (2010) 083.
\bibitem{11}
C. E. Yaguna, JHEP {\bf 1108}  (2011) 060.
\bibitem{12}
R. Coimbra, M. O. P. Sampaio and R. Santos, Eur. Phys. J. C {\bf 73}
(2013) 2428.
\bibitem{13}
L. Feng, A. S. Profumo and B. L. Ubaldi, JHEP {\bf 03} (2015) 045.
\bibitem{14}
J. R. Espinosa and M. Quiros, Phys. Lett. B{\bf 305} (1993) 98.
\bibitem{15}
J. Choi and R. R. Volkas, Phys. Lett. B{\bf 317} (1993) 385.
\bibitem{16}
S. Profumo, M. J. Ramsey-Musolf, G. Shaughnessy, JHEP {\bf 0708}
(2007) 010.
\bibitem{17}
A. Noble and M. Perelstein, Phys. Rev. D {\bf 78} (2008) 063518.
\bibitem{18}
A. Ashoorioon and T. Konstandin, JHEP {\bf 0907} (2009) 086.
\bibitem{19}
S. Das, P. J. Fox, A. Kumar and N. Weiner, JHEP {\bf 1011} (2010)
108.
\bibitem{20}
J. R. Espinosa, T. Konstandin, and F. Riva, Nucl. Phys. B {\bf 854}
(2012) 592.
\bibitem{21}
T. Alanne, K. Tuominen, and V. Vaskonen, Nucl. Phys. B {\bf 889}
(2014) 692.
\bibitem{22}
D. Curtin, P. Meade and C. T. Yu, JHEP {\bf 1411} (2014) 127.
\bibitem{23}
S. Profumo, M J. Ramsey-Musolf, C. L. Wainwright and P. Winslow,
Phys. Rev. D {\bf 91} (2015) 035018.
\bibitem{24}
J. M. Cline, K. Kainulainen, P. Scott and C. Weniger, Phys. Rev. D
{\bf 88}  (2013) 055025.
\bibitem{25}
J. M. Cline, K. Kainulainen, JCAP {\bf 1301} (2013) 012.
\bibitem{26}
J. R. Espinosa and M. Quiros, Phys. Rev. D {\bf 76} (2007) 07600.
\bibitem{27}
J. R. Espinosa, T. Konstandin, J. M. No and M. Quiros, Phys. Rev. D
{\bf 78} (2008) 123528.
\bibitem{28}
A. Drozd, B. Grzadkowski and J. Wudka, Acta Phys. Polon. B {\bf 42}
(2011) 2255.
\bibitem{29}
A. Abada, S. Nasri and D. Ghaffor, Phys. Rev. D {\bf 83} (2011)
095021.
\bibitem{30}
A. Abada and S. Nasri, Phys. Rev. D {\bf 85} (2012) 075009.
\bibitem{31}
A. Drozd, B. Grzadkowski and J. Wudka, JHEP {\bf 1204} (2012) 006.
\bibitem{32}
A. Ahriche and S. Nasri, Phys. Rev. D {\bf 85} (2012) 093007.
\bibitem{33}
A. Ahriche, A. Arhrib and S. Nasri, JHEP {\bf 02} (2014) 042.
\bibitem{34}
A. Tofighi, O. N. Ghodsi and M. Saeedhoseini, Phys. Lett. B {\bf
748} (2015) 208.
\bibitem{35}
K. P. Modak, D. Majumdar and S. Rakshit, JCAP {\bf 1503} (2015) 011.
\bibitem{36}
R. Jinno, K. Nakayama and M. Takimoto, Phys. Rev. D {\bf 93}  (2016)
045024.
\bibitem{37}
 J. M. Cline and P.-A. Lemieux, Phys.Rev. D {\bf 55} (1997) 3873.
\bibitem{38}
L. Fromme, S. J. Huber, and M. Seniuch, JHEP {\bf 0611} (2006) 038.
\bibitem{39}
S. J. Huber and M. G. Schmidt, Eur. Phys. J. C {\bf 10} (1999) 473.
\bibitem{40}
K. Funakubo and E. Senaha, Phys. Rev. D {\bf 79} (2009) 115024.
\bibitem{41}
D. J. H. Chung and A. J. Long, Phys. Rev. D {\bf 81}  (2010) 123531.
\bibitem{42}
J. Kozaczuk, S. Profumo, L. S. Haskins and C. L. Weinwright, JHEP
{\bf 1501} (2015) 144.
\bibitem{43}
C. Balazs, A. Mazumdar, E. Pukartas and G. White, JHEP {\bf 1401}
(2014) 073.
\bibitem{44}
W. Huang, Z. Kang, J. Shu and J. M. Yang, Phys. Rev. D {\bf 91}
(2015) 025006.
\bibitem{45}
C. -W. Chiang and E. Senaha, JHEP {\bf 1006} (2010) 030.
\bibitem{46}
D. J. Chung, A. J. Long, and L.-T. Wang, Phys. Rev. D {\bf 87}
(2013) 023509.
\bibitem{47}
L. Dolan and R. Jackiw, Phys. Rev. D {\bf 9} (1974) 3320.
\bibitem{48}
E. W. Kolb and M. S. Turner, \textit {The Early Universe}, Front.
Phys. {\bf 69} (1990) 1–547.
\bibitem{49}
J. I. Kapusta, \textit{Finite emperature field theory}, Cambridge
University Press, (1989).
\bibitem{50}
M. Le-Bellac, \textit{Thermal field theory}, Cambridge University
Press, (2000).
\bibitem{51}
 A. Kosowsky, A. Mack and T. Kahniashvili, Phys. Rev. D {\bf66} (2002) 024030.
\bibitem{52}
 A. Kosowsky, M. S. Turner and R. Watkins, Phys. Rev. D {\bf 45} (1992) 4514.
\bibitem{53}
 M. Kamionkowski, A. Kosowsky and M. S. Turner, Phys. Rev. D {\bf 49} (1994)
 2837.
 \bibitem{54}
 C. Caprini and R. Durrer, Phys. Rev. D {\bf74}  (2006) 063521.
\bibitem{55}
L. Leitao and A. Megevand, arxiv:1512.08962.
\bibitem{56}
M. Artymowski, M. Lewicki and J. D. Wells, arxiv:1609.07143.
\bibitem{57}
M. Kakizaki, S. Kanemura and T. Matsui, Phys. Rev. D {\bf 92} (2015)
115007.
\bibitem{58}
P. Huang, A. J. Long, L.-T. Wang, Phys. Rev. D {\bf 95} (2016)
075008.
\bibitem{59}
C. Balazs, A. Fowlie, A. Mazumdar and G. White, Phys. Rev. D {\bf
95} (2017) 043505.
\bibitem{60}
A. Katz and M. Perelstein, JHEP {\bf 1407} (2014) 108.
\bibitem{61}
F. Sannino and J. Virkaja$\ddot{a}$rvi, Phys. Rev. D {\bf 92} (2015)
045015.
\bibitem{62}
Z. -W. Wang, T. G. Steele, T. Hanif and R. B. Mann, JHEP {\bf
08}(2016) 065.
\bibitem{63}
M. Lewicki, T. Rindler-Daller and J. D. Wells, JHEP {\bf 06} (2016)
055.
\bibitem{64}
A. V. Kotwal, M. J. Ramsey-Musolf, J. M. No and P. Winslow, Phys.
Rev. D {\bf 94} (2016) 035022.
\bibitem{65}
I. Baldes, T. Konstandin and J. Servant, arxiv:1604.04526.
\bibitem{66}
T. Tenkanen, K. Tuominen and V. Vaskonen, arxiv:1606.06063.
\bibitem{67}
M. Chala, G. Nardinib and I. Sobolevc, Phys. Rev. D {\bf 94} (2016)
055006.
\bibitem{68}
V. A. Kuzmin, V. A. Rubakov and M. E. Shaposhnikov, Phys. Lett. B
{\bf 155} (1985) 36.
\bibitem{69}
D. E. Morrissey and M. J. Ramsey-Musolf, New J. Phys. {\bf 14}
(2012) 125003.
\bibitem{70}
S. Coleman, \textit{Aspects of symmetry: Selected Erice lectures},
Cambridge University Press, Cambridge, (1988).
\bibitem{71}
A. Zee, \textit{Quantum field theory in a nutshell}, Princeton
University Press, Princeton, (2010).
\bibitem{72}
M. Carena, A. Megevand, M. Quiros and C. E. M. Wagner, Nucl. Phys. B
{\bf 716} (2005) 319.
\bibitem{73}
P. B. Arnold and O. Espinosa, Phys. Rev. D {\bf 47} (1993) 3546.
\bibitem{74}
Z. Fodor and A. Hebecker, Nucl. Phys. B {\bf 432} (1994) 127.
\bibitem{75}
D. Curtin, P. Meade and H. Ramani, arxiv:1612.00466.
\bibitem{76}
 G. $'t$ Hooft, Phys. Rev. Lett. {\bf 37} (1976) 8.
\bibitem{77}
G. $'t$ Hooft, Phys. Rev. D {\bf 14} (1976) 3432.
\bibitem{78}
N. S. Manton, Phys. Rev. D {\bf 28} (1983) 2019.
\bibitem{79}
S. Kanemura, M. Kikuchia and K. Yagyub, Nucl. Phys. B {\bf 917}
(2017) 154.
\bibitem{80}
T. Robens and T. Stefaniak, Eur. Phys. J. C {\bf 76} (2016) 268.
\bibitem{81}
K. Cheung, P. Ko, J. S. Lee and P. Y. Tseng, JHEP {\bf 10} (2015)
057.
\bibitem{82}
P. Bechtle, S. Heinemeyer, O. St{\aa}l, T. Stefaniak, and G.
Weiglein, JHEP {\bf 11} (2014) 039.
\bibitem{83}
Virgo, LIGO Scientific collaboration, B. P. Abbott et al., Phys.
Rev. Lett. {\bf 116} (2016) 061102.
\bibitem{84}
 C. Caprini, et al., JCAP {\bf 1604}
 (2016) 001.
 \bibitem{85}
V. Vaskonen, arxiv:1611.02073.
\bibitem{86}
 K. Kannike, Eur. Phys. J. C {\bf 72}, 2093 (2012) 2093.
\bibitem{87}
J. Chakrabortty, P. Konar and T. Mondal, arXiv:1311.5666.
\end{thebibliography}
\end{document}